\def\papertitle{Compiling Signal Processing Code embedded in Haskell via LLVM}
\def\paperauthorA{Henning Thielemann}

\documentclass[twoside,a4paper]{article}
\usepackage{dafx_10}
\usepackage{amsmath,amssymb,amsfonts,amsthm}
\usepackage{euscript}
\usepackage[latin1]{inputenc}
\usepackage[T1]{fontenc}
\usepackage{ifpdf}

\usepackage[english]{babel}
\usepackage{caption}
\usepackage{subfig, color}

\ninept

\usepackage{times}

\ifpdf 
  \usepackage[pdftex,
    pdftitle={\papertitle},
    pdfauthor={\paperauthorA},
    colorlinks=false, 
    bookmarksnumbered, 
    pdfstartview=XYZ 
  ]{hyperref}
  \usepackage[pdftex]{graphicx}
  \usepackage[figure,table]{hypcap}
\else 
  \usepackage[dvips]{epsfig,graphicx}
  \usepackage[dvips,
    colorlinks=false, 
    bookmarksnumbered, 
    pdfstartview=XYZ 
  ]{hyperref}
  \usepackage[figure,table]{hypcap}
\fi

\title{\papertitle}

\usepackage{tikz}
\usepackage{comment}
\usepackage{mathpaper}
\usepackage{IEEEtrantools}

\newcommand\reviewernote[1]{}

\newcommand\temporarynote[1]{\textcolor{red}{#1}}
\newcommand\code[1]{\texttt{#1}}

\newcommand\haskell{\texttt{Haskell}}
\newcommand\abbreviation[2]{\hypertarget{abbrev:#1}{#2} (#1)}
\newcommand\abbr[1]{\hyperlink{abbrev:#1}{#1}}
\newcommand\shiftup{\mathbin{>\!\!\!>}}  
\DeclareMathOperator{\fraction}{frac}

\newcommand\figcaption[2]{\caption{\textit{#2}}\figlabel{#1}}
\newcommand\tabcaption[2]{\caption{\textit{#2}}\tablabel{#1}}

\affiliation{
\paperauthorA}
{\href{http://www.informatik.uni-halle.de}{Institut f\"{u}r Informatik,
Martin-Luther-Universit\"{a}t Halle-Wittenberg, Germany} \\
{\tt \href{mailto:henning.thielemann@informatik.uni-halle.de}{henning.thielemann@informatik.uni-halle.de}}
}

\begin{document}
\ifpdf 
  \DeclareGraphicsExtensions{.png,.jpg,.pdf}
\else  
  \DeclareGraphicsExtensions{.eps}
\fi

\maketitle

\begin{abstract}
We discuss a programming language for real-time audio signal processing
that is embedded in the functional language Haskell
and uses the Low-Level Virtual Machine as back-end.
With that framework we can code with the comfort and type safety of Haskell
while achieving maximum efficiency of fast inner loops and full vectorisation.
This way Haskell becomes a valuable alternative
to special purpose signal processing languages.
\reviewernote{
For the presentation I plan to show
some examples of interactive compilation and
real-time control of signal processes.
}
\end{abstract}

\section{Introduction}

\reviewernote{
The organisation of this article follows the advises of
\href%
{http://research.microsoft.com/en-us/um/people/simonpj/papers/giving-a-talk/giving-a-talk.htm}%
{http://research.microsoft.com/en-us/um/people/simonpj/papers/giving-a-talk/giving-a-talk.htm}.
}
Given a data flow diagram as in \figref{goal-dataflow}
we want to generate an executable machine program as in \figref{goal-assembly}.
First the diagram must be translated
to something that is more accessible by a machine.
Since we can translate data flows almost literally to function expressions
we choose a functional programming language as the target language,
here \haskell{} \cite{peyton-jones1998haskell}.
The result can be seen in \figref{goal-haskell}.
This translation must be done manually
but in future it could also be supported by a flow diagram editor
like PureData \cite{puckette1996puredata} or Gems \cite{evans2007gemcutter}.
The second step is to translate the function expression
to a machine oriented presentation.
This is the main concern of our paper.

Since we represent signals as sequences of numbers,
signal processing algorithms are usually loops
that process these numbers one after another.
Thus our goal is to generate efficient loop bodies
as in \figref{goal-llvm}
from a functional signal processing representation.
We have chosen the \abbreviation{LLVM}{Low-Level Virtual-Machine}
\cite{lattner2004llvm}
for the loop description,
because \abbr{LLVM} %
provides a universal representation
for machine languages of a wide range of processors.
The \abbr{LLVM} library is responsible for the third step,
namely the translation of portable virtual machine code
to actual machine code of the host processor.

Our contributions are
\begin{itemize}
\item a representation of an \abbr{LLVM} loop body
that can be treated like a signal, described in \secref{signal-generator},
\item a way to describe causal signal processes
which is the dominant kind of signal transformations
in real-time audio processing
and which allows us to cope efficiently with multiple uses of outputs
and with feedback of even small delays, guaranteed deadlock free,
developed in \secref{causal-process},
\item a handling of internal filter parameters
in a way that is much more flexible
than traditional control rate/sample rate schemes,
presented in \secref{internal-parameter},
\item support for the vector units of modern processors
both for non-recursive and recursive signal processes
as derived in \secref{vectorisation},
\item a method for compiling a signal processing algorithm once
and run it with different parameters
as shown in \secref{parameter}
\item and a speed comparison with well established signal processing packages
in \secref{benchmark}.
\end{itemize}

\tikzstyle{process}=[rectangle,draw=black!50,fill=black!3,thick]

\begin{figure}
\begin{tikzpicture}[inner sep=2ex]
\pgfsetarrowsend{latex}
  \node[process] (oscillator)  at (0,0) {oscillator};
  \node[process] (exponential) at (0,2) {exponential};
  \node[process] (amplifier)   at (4,0) {amplifier};
  \draw (oscillator.east)  -- node[above,pos=0.20](osc-out){} (amplifier.west);
  \draw (exponential.east) -| node[above,pos=0.05](exp-out){} (amplifier.north);
  \draw (amplifier.east)   -- node[above,near start](amp-out){} (6,0);

\pgfsetarrowsend{}
  \begin{scope}[shift={(osc-out)},scale=0.3]
  \draw[->,black!30] (0,0) -- (5.5,0);
  \draw[samples=51,domain=0:5] plot[id=saw]  function{1-2*(x-floor(x))};
  \end{scope}

  \begin{scope}[shift={(exp-out)},scale=0.3]
  \draw[->,black!30] (0,0) -- (5.5,0);
  \draw[samples=51,domain=0:5] plot[id=exp]  function{0.5**x};
  \end{scope}

  \begin{scope}[shift={(amp-out)},scale=0.3]
  \draw[->,black!30] (0,0) -- (5.5,0);
  \draw[samples=51,domain=0:5] plot[id=ping] function{0.5**x*(1-2*(x-floor(x)))};
  \end{scope}
\end{tikzpicture}
%
\figcaption{goal-dataflow}{
Data flow for creation of a very simple percussive sound
}
\end{figure}

\begin{figure}
\begin{verbatim}
amplify
   (exponential halfLife amp)
   (osci Wave.saw phase freq)
\end{verbatim}
\vspace{-\baselineskip}
\figcaption{goal-haskell}{
Functional expression for the diagram in \figref{goal-dataflow}
}
\end{figure}

\begin{figure}
\begin{verbatim}
define generate
   (phase, freq, amp, decay, ptr, counter) {
  goto enter

loop:
  y0 := mul 2, phase   ; phase \in [0,1]
  y1 := sub 1, y0      ; saw = 1-2*phase
  y2 := mul amp, y1    ; out = amp * saw
  store y2, ptr        ; write to buffer

  amp := mul amp, decay
                       ; update amp
  phaseTmp := add phase, freq
                       ; update phase
  phase := fraction phaseTmp
                       ; wrap phase to [0,1]

  counter := sub counter, 1
  ptr := add ptr, sizeof(y2)

enter:
  if counter>0 then goto loop
}
\end{verbatim}
\vspace{-\baselineskip}
\figcaption{goal-llvm}{
Simplified LLVM assembly code
to be generated from the function expression in \figref{goal-haskell}.
In our example scalar and vectorised loop are the same.
}
\end{figure}

\section{Background}
\seclabel{background}

We want to generate \abbr{LLVM} code from a signal processing algorithm
written in a declarative way.
We like to write code close to a data flow diagram
and the functional paradigm seems to be appropriate.

\begin{figure}
\begin{verbatim}
  # (phase, phase+freq, phase+2*freq, ...)
  movaps    phases, %xmm0
  # (amp, amp/2^(-1/halfLife), ...)
  movaps    amps,   %xmm1
  # (4*freq, 4*freq, 4*freq, ...)
  movaps    freqs,  %xmm2
  # (2^(-4/halfLife), 2^(-4/halfLife), ...)
  movaps    decays, %xmm3
  movaps    ones,   %xmm4 # (1,1,1,1)
  jmp       enter
loop:
  movaps    %xmm3, %xmm5
  mulps     %xmm1, %xmm5  # amp*decays
  movaps    %xmm2, %xmm6
  addps     %xmm0, %xmm6  # t' := t+freq
  mulps     twos,  %xmm0  # 2*t
  movaps    %xmm4, %xmm7
  subps     %xmm0, %xmm7  # 1-2*t
  mulps     %xmm1, %xmm7  # amp*(1-2*t)
  movaps    %xmm7, (%ecx) # store
  addl      $16, %ecx     # next packet
  cvttps2dq %xmm6, %xmm0
  cvtdq2ps  %xmm0, %xmm0  # truncate t'
  subps     %xmm0, %xmm6  # fraction t'
  movaps    %xmm5, %xmm1
  movaps    %xmm6, %xmm0
  decl      %edx          # counter
enter:
  testl     %edx, %edx
  jne       loop
\end{verbatim}
\vspace{-\baselineskip}
\figcaption{goal-assembly}{
Intel x86 assembly loop for \figref{goal-llvm}
using SSE vector instructions.
}
\end{figure}

We could design a new language specifically for this purpose,
but we risk the introduction of design flaws.
We could use an existing signal processing language,
but we are afraid that it does not scale well
to applications other than signal processing.
Alternatively we can resort to an existing
general purpose functional programming language or a subset of it,
and write a compiler with optimisations adapted to signal processing needs.
But writing a compiler for any modern ``real-world'' programming language
is a task of several years, if not decades.
A compiler for a subset of an existing language however
would make it hard to interact with existing libraries.
So we can still tune an existing compiler for an existing language,
but given the complexity of modern languages and their respective compilers
this is still a big effort.
It might turn out that a change that is useful for signal processing
kills performance for another application.

A much quicker way to adapt a language to a special purpose
is the \abbreviation{EDSL}{Embedded Domain Specific Language} approach
\cite{landin1966dsl}.
In this terminology ``embedded'' means
that the domain specific (or ``special purpose'') language
is actually not an entirely new language,
but a way to express domain specific issues
using corresponding constructs and checks of the host language.
For example, writing an SQL command as string literal in Java
and sending it to a database, is not an \abbr{EDSL}.
In contrast to that, Hibernate \cite{elliott2004hibernate} is an \abbr{EDSL},
because it makes database table rows look like ordinary Java objects
and it makes the use of foreign keys safe and comfortable
by making foreign references look like Java references.

In the same way we want to cope with signal processing in \haskell{}.
In the expression
\begin{verbatim}
amplify
   (exponential halfLife amp)
   (osci Wave.saw phase freq)
\end{verbatim}
the call to \code{osci} shall not produce a signal,
but instead it shall generate \abbr{LLVM} code
that becomes part of a signal generation loop later.
In the same way \code{amplify} assembles the code parts
produced by \code{exponential} and \code{osci}
and defines the product of their results as its own result.
In the end every such signal expression
is actually a high-level \abbr{LLVM} macro
and finally, we pass it to a driver function that compiles and runs the code.
Where Hibernate converts Java expressions to SQL queries,
sends them to a database
and then converts the database answers back to Java objects,
we convert \haskell{} expressions to LLVM bitcode,
send it to the LLVM \abbreviation{JIT}{Just-In-Time} compiler
and then execute the resulting code.
We can freely exchange signal data between pure \haskell{} code
and LLVM generated code.

The \abbr{EDSL} approach is very popular among \haskell{} programmers.
For instance interfaces to the CSound signal processing language
\cite{hudak1996haskore}
and the real-time software synthesiser SuperCollider
\cite{drape2009hsc3-0.7}
are written this way.
This popularity can certainly be attributed
to the concise style of writing \haskell{} expressions
and to the ease of overloading number literals and arithmetic operators.
We shall note that the \abbr{EDSL} method has its own shortcomings,
most notably the \keyword{sharing problem}
that we tackle in \secref{causal-process}.

In \cite{thielemann2004haskellsignal} we have argued extensively,
why we think that \haskell{} is a good choice for signal processing.
Summarised, the key features for us are
polymorphic but strong static typing and lazy evaluation
\cite{hughes1984fpmatter}. 
Strong typing means that we have a wide range of types
that the compiler can distinguish between.
This way we can represent a trigger or gate signal
by a sequence of boolean values (type \code{Bool})
and this cannot be accidentally mixed up with a PCM signal
(sample type \code{Int8}),
although both types may be represented by bytes internally.
We can also represent internal parameters of signal processes
by opaque types that can be stored by the user but cannot be manipulated
(cf.\ \secref{internal-parameter}).
Polymorphic typing means that we can write a generic algorithm
that can be applied
to single precision or double precision floating point numbers,
to fixed point numbers or complex numbers.
Static typing means that the \haskell{} compiler can check
that everything fits together
when compiling a program or parts of it.
Lazy evaluation means that we can transform audio data
as it becomes available
while programming in a style that treats those streams
as if they were available at once.

The target language of our embedded compiler is \abbr{LLVM}.
It differs from CSound and SuperCollider
in that \abbr{LLVM} is not a signal processing language.
It is a high-level assembler and we have to write
the core signal processing building blocks ourselves.
However, once this is done assembling those blocks
is as simple as writing CSound or SuperCollider/SCLang programs.
We could have chosen a concrete machine language as target,
but \abbr{LLVM} does a much better job for us:
It generates machine code for many different processors,
thus it can be considered a portable assembler.
It also supports the vector units of modern processors (\secref{vectorisation}) and
target dependent instructions (\keyword{intrinsics})
and provides us with a large set of low-level to high-level optimisations,
that we can even select and arrange individually.
We can run \abbr{LLVM} code immediately from our \haskell{} programs
(\abbr{JIT}),
but we can also write \abbr{LLVM} bitcode files
for debugging or external usage.

\section{Implementation}

We are now going to discuss the design of our implementation
\cite{thielemann2010synthesizer,thielemann2010synthillvm-darcs}.

\subsection{Signal generator}
\seclabel{signal-generator}

In our design a signal is a sequence of sample values
and a signal generator is a state transition system
that ships a single sample per request while updating the state.
E.g.\ the state of an exponential curve is the current amplitude
and on demand it returns the current amplitude as result
while decreasing the amplitude state by a constant factor.
In the same way an oscillator uses the phase as internal state.
Per request it applies a wave function on the phase
and delivers the resulting value as current sample.
Additionally it increases the phase by the oscillator frequency
and wraps around the result to the interval $[0,1)$.
This design is much inspired by \cite{coutts2007streamfusion}.

According to this model
we define an \abbr{LLVM} signal generator in \haskell{} essentially as
a pair of an initial state
and a function that returns a tuple
containing a flag showing whether there are more samples to come,
the generated sample and the updated state.
\begin{verbatim}
type Generator a =
   (state,
    state -> Code (Value Bool, (a, state)))
\end{verbatim}
The lower-case identifiers are \keyword{type variables}
that can be instantiated with actual types.
The variable \code{a} is for the sample type
and \code{state} for the internal state of the signal generator.
Since \code{Generator} is not really a signal
but a description for \abbr{LLVM} code,
the sample type cannot be just a \haskell{} number type
like \code{Float} or \code{Double}.
Instead it must be the type for one of \abbr{LLVM}'s virtual registers,
namely \code{Value Float} or \code{Value Double}, respectively.
The types \code{Value} and \code{Code}
are imported from a \haskell{} interface to \abbr{LLVM}
\cite{osullivan2010llvm-0.7.1.1}.

The type parameter is not restricted in any way,
thus we can implement a generator of type
\code{Generator (Value Float, Value Float)}
for a stereo signal generator or
\code{Generator (Value Bool, Value Float)}
for a gate signal and a continuous signal
that are generated synchronously.
We do not worry about a layout in memory of an according signal at this point,
since it may be just an interim signal
that is never written to memory.
E.g.\ the latter of the two types just says, that the generated samples
for every call to the generator can be found in two virtual registers,
where one register holds a boolean and the other one a floating point number.

We like to complement this general description
with the simple example of an exponential curve generator.
\begin{verbatim}
exponential ::
   Float -> Float -> Generator (Value Float)
exponential halfLife amp =
   (valueOf amp,
    \y0 -> do
      y1 <- mul y0 (valueOf
                (2**(-1/halfLife) :: Float))
      return (valueOf True, (y0, y1)))
\end{verbatim}
For simplification we use the fixed type \code{Float}
but in the real implementation the type is flexible.
The implementation is the same,
only the real type of \code{exponential} is considerably more complicated
because of many constraints to the type parameters.

The function \code{valueOf} 
makes a \haskell{} value available as constant in \abbr{LLVM} code.
Thus the power computation with \code{**} in the \code{mul} instruction
is done by \haskell{} and then implanted into the \abbr{LLVM} code.
This also implies that the power is computed only once.
The whole transition function, that is the second element of the pair,
is a \keyword{lambda expression}, also known as \keyword{anonymous function}.
It starts with a backslash and its argument \code{y0}
which identifies the virtual register
that holds the current internal state.
It returns always \code{True} because it never terminates
and it returns the current amplitude \code{y0} as current sample
and the updated amplitude computed by a multiplication
to be found in the register identified by \code{y1}.

We have seen how basic signal generators work,
however, signal processing consists largely of transforming signals.
In our framework a signal transformation
is actually a generator transformation.
That is we take apart given generators
and build something new from them.
For example the controlled amplifier
dissects the envelope generator and the input generator
and assembles a generator for the amplified signal.
\begin{verbatim}
amplify ::
   Generator (Value Float) ->
   Generator (Value Float) ->
   Generator (Value Float)
amplify (envInit, envTrans)
        (inInit,  inTrans) =
   ((envInit, inInit),
    (\(e0,i0) -> do
       (eCont,(ev,e1)) <- envTrans e0
       (iCont,(iv,i1)) <- inTrans  i0
       y    <- mul ev iv
       cont <- and eCont iCont
       return (cont, (y, (e1,i1)))))
\end{verbatim}

So far our signals only exist as \abbr{LLVM} code,
but computing actual data is straightforward:
\begin{verbatim}
render ::
   Generator (Value Float) ->
   Value Word32 -> Value (Ptr Float) ->
   Code (Value Word32)
render (start, next) size ptr = do
   (pos,_) <- arrayLoop size ptr start $
      \ ptri s0 -> do
         (cont,(y,s1)) <- next s0
         ifThen cont () (store y ptri)
         return (cont, s1)
   ret pos
\end{verbatim}
The ugly branching that is typical for assembly languages
including that of \abbr{LLVM}
is hidden in our custom functions \code{arrayLoop} and \code{ifThen}.
\haskell{} makes a nice job as macro assembler.
Again, we only present the most simple case here.
The alternative to filling a single buffer with signal data is
to fill a sequence of chunks that are created on demand.
This is called \keyword{lazy evaluation}
and one of the central features of \haskell{}.

At this point, we might wonder
whether the presented model of signal generators is general enough
to match all kinds of signals that can appear in real applications.
The answer is yes, since given a signal
there is a generator that emits that signal.
We simply write the signal to a buffer
and then use a signal generator
that manages a pointer into this buffer as internal state.
This generator has a real-world use when reading a signal from a file.
We see that our model of signal generators
does not impose a restriction on the kind of signals,
but it well restricts the access to the generated data:
We can only traverse from the beginning to the end of the signal
without skipping any value.
This is however intended since we want to play the signals in real-time.

\subsection{Causal Processes}
\seclabel{causal-process}

While the above approach of treating signal transformations
as signal generator transformations is very general
it can be inefficient.
For example for a signal generator \code{x}
the expression \code{mix x x}
does not mean that the signal represented by \code{x} is computed once
and then mixed with itself.
Instead, the mixer runs the signal generator \code{x} twice
and adds the results of both instances.
I like to call that the \keyword{sharing problem}.
It is inherent to all DSLs that are embedded into a purely functional language,
since in those languages objects have no identity,
i.e.\ you cannot obtain an object's address in memory.
The sharing problem also occurs if we process
the components of a multi-output signal process individually,
for instance the channels of a stereo signal
or the lowpass, bandpass, highpass components of a state variable filter.
E.g.\ for delaying the right channel of a stereo signal
we have to write \code{stereo (left x) (delay (right x))}
and we run into the sharing problem, again.

We see two ways out:
The first one is relying on \abbr{LLVM}'s optimiser
to remove the duplicate code.
However this may fail
since \abbr{LLVM} cannot remove duplicate code
if it relies on seemingly independent states,
on interaction with memory
or even on interaction with the outside world.
Another drawback is that the temporarily generated code may grow exponentially
compared to the code written by the user.
E.g.\ in
\begin{verbatim}
let y = mix x x
    z = mix y y
in  mix z z
\end{verbatim}
the generator \code{x} is run eight times.

The second way out is to store the results of a generator
and share the storage amongst all users of the generator.
We can do this by rendering the signal to a lazy list,
or preferably to a lazily generated list of chunks for higher performance.
This approach is a solution to the general case
and it would also work if there are signal processes involved
that shrink the time line,
like in \code{mix x (timeShrink x)}.

While this works in the general case
there are many cases where it is not satisfying.
Especially in the example \code{mix x x}
we do not really need to store the result of \code{x} anywhere
as it is consumed immediately by the mixer.
Storing the result is at least inefficient
in case of a plain \haskell{} singly linked list
and even introduces higher latency in case of a chunk list.

So what is the key difference between
\code{mix x x} and \code{mix x (timeShrink x)}?
It is certainly that in the first case data is processed in a synchronous way
thus it can be consumed (mixed) as it is produced (generated by \code{x}).
However, the approach of signal transformation by signal generator transformation
cannot model this behaviour.
When considering the expression \code{mix x (f x)}
we have no idea whether \code{f}
maintains the ``speed'' of its argument generator.
That is, we need a way to express that \code{f} emits data synchronously to its input.
For instance we could define
\begin{verbatim}
type Map a b = a -> Code b
\end{verbatim}
that represents a signal transformation
of type \code{Generator a -> Generator b}.
It could be applied to a signal generator by a function \code{apply} with type
\begin{verbatim}
Map a b -> Generator a -> Generator b
\end{verbatim}
and where we would have written \code{f x} before,
we would write \code{apply f x} instead.

It turns out that \code{Map} is too restrictive.
Our signal process would stay synchronous
if we allow a running state as in a recursive filter
and if we allow termination of the signal process
before the end of the input signal
as in the \haskell{} list function \code{take}.
Thus, what we actually use is a definition that boils down to
\begin{verbatim}
type Causal a b =
   (state, (a, state) ->
           Code (Value Bool, (b, state)))  .
\end{verbatim}
With this type we can model all kinds of causal processes,
that is processes where every output sample
depends exclusively on the current and past input samples.
The \code{take} function may serve as an example
for a causal process with termination.
\begin{verbatim}
take :: Int -> Causal a a
take n =
  (valueOf n,
   \(a,toDo) -> do
      cont <- icmp IntULT (valueOf 0) toDo
      stillToDo <- sub toDo (valueOf 1)
      return (cont, (a, stillToDo)))
\end{verbatim}

The function \code{apply} for applying a causal process to a signal generator
has the signature
\begin{verbatim}
apply ::
   Causal a b -> Generator a -> Generator b
\end{verbatim}
and its implementation is straightforward.
The function is necessary to do something useful with causal processes,
but it loses the causality property.
For sharing we want to make use of facts like
that the serial composition of causal processes is causal, too,
but if we have to express the serial composition
of processes \code{f} and \code{g} by \code{apply f (apply g x)},
then we cannot make use of such laws.
The solution is to combine processes with processes
rather than transformations with signals.
E.g.\ with \verb|>>>| denoting the serial composition
we can state that \verb|g >>> f| is a causal process.

In the base \haskell{} libraries there is already the \code{Arrow} abstraction
that was developed for the design of integrated circuits in the Lava project,
but it proved to be useful for many other applications.
The \code{Arrow} type class provides a generalisation
of plain \haskell{} functions.
For making \code{Causal} an instance of \code{Arrow}
we must provide the following minimal set of methods
and warrant the validity of the arrow laws \cite{hughes2000arrows}.
\begin{verbatim}
arr :: (a -> b) -> Causal a b
(>>>) ::
  Causal a b -> Causal b c -> Causal a c
first :: Causal a b -> Causal (a,c) (b,c)
\end{verbatim}
The infix operator \verb|>>>| implements (serial) function composition,
the function \code{first} allows for parallel composition,
and the function \code{arr} generates stateless transformations
including rearrangement of tuples as needed by \code{first}.
It turns out that all of these \keyword{combinators} maintain causality.
They allow us to express all kinds of causal processes without feedback.
If \code{f} and \code{mix} are causal processes,
then we can translate the former \code{mix x (f x)} to
\begin{verbatim}
arr (\x -> (x,x)) >>> second f >>> mix
where second p = arr swap >>> p >>> arr swap
      swap (a,b) = (b,a)  .
\end{verbatim}

For implementation of feedback
we need only one other combinator, namely \code{loop}.
\begin{verbatim}
loop :: c -> Causal(a,c) (b,c) -> Causal a b
\end{verbatim}
The function \code{loop} feeds the output of type \code{c}
of a process back to its input channel of the same type.
In contrast to the \code{loop} method of the standard \code{ArrowLoop} class
we must delay the value by one sample
and thus need an initial value of type \code{c} for the feedback signal.
Because of the way \code{loop} is designed it cannot run into deadlocks.
In general deadlocks can occur whenever a signal processor runs ahead of time,
that is it requires future input data
in order to compute current output data.
Our notion of a causal process excludes this danger.

In fact, feedback can be considered another instance of the sharing problem
and \code{loop} is its solution.
For instance if we want to compute a comb filter
for input signal \code{x} and output signal \code{y}
then the most elegant solution in \haskell{}
is to represent \code{x} and \code{y} by lists and write the equation
\code{let y = x + delay y in y}
which can be solved lazily by the \haskell{} runtime system.
In contrast to that if \code{x} and \code{y} are signal generators
this would mean to produce infinitely large code since it holds
\begin{verbatim}
y = x + delay y
  = x + delay (x + delay y)
  = x + delay (x + delay (x + delay y)) ...
\end{verbatim}
With \code{loop} however we can share the output signal \code{y}
with its occurrences on the right hand side.
Therefore, the code would be
\begin{verbatim}
y = apply (arr mixFanout >>> second delay) x
        where mixFanout (a,b) = (a+b,a+b)  .
\end{verbatim}

Since the use of arrow combinators is somehow less intuitive
than regular function application and \haskell{}'s recursive \code{let} syntax,
there is a preprocessor that translates a special arrow syntax
into the above combinators.
Further on there is a nice abstraction of causal processes,
namely commutative causal arrows \cite{liu2009cca}.

We like to note that we can even express signal processes
that are causal with respect to one input
and non-causal with respect to another one.
E.g.\ frequency modulation
is causal with respect to the frequency control
but non-causal with respect to the input signal.
This can be expressed by the type
\begin{verbatim}
freqMod :: Generator (Value a) ->
           Causal (Value a) (Value a)  .
\end{verbatim}

In retrospect, our causal process data type
looks very much like the signal generator type.
It just adds a parameter to the transition function.
Vice versa the signal generator data type could be replaced
by a causal process with no input channel.
We could express this by
\begin{verbatim}
type Generator a = Causal () a
\end{verbatim}
where \code{()} is a nullary tuple.
However for clarity reasons we keep \code{Generator} and \code{Causal} apart.

\subsection{Internal parameters}
\seclabel{internal-parameter}

It is a common problem in signal processing
that recursive filters \cite{hamming1989digitalfilter} are cheap in execution
but computation of their internal parameters
(mainly feedback coefficients) is expensive.
A popular solution to this problem is to compute the filter parameters
at a lower sampling rate \cite{vercoe2004csound,mccartney1996supercollider}.
Usually, the filter implementations
hide the existence of internal parameters
and thus they have to cope with the different sampling rates themselves.

In this project we choose a more modular way.
We make the filter parameters explicit but opaque
and split the filtering process
into generation of filter parameters,
filter parameter resampling and actual filtering.
Static typing asserts that filter parameters
can only be used with the respective filters.

This approach has several advantages:
\begin{itemize}
\item
A filter only has to treat inputs of the same sampling rate.
We do not have to duplicate the code for coping with input
at rates different from the sample rate.
\item We can provide different ways of specifying filter parameters,
e.g.\ the resonance of a lowpass filter can be controlled
either by the slope or by the amplification of the resonant frequency.
\item We can use different control rates in the same program.
\item We can even adapt the speed of filter parameter generation
to the speed of changes in the control signal.
\item For a sinusoidal controlled filter sweep we can setup a table
of filter parameters for logarithmically equally spaced cutoff frequencies
and ship this table at varying rates according to arcus sine.
\item Classical handling of control rate filter parameter computation
can be considered as resampling of filter parameters with constant interpolation.
If there is only a small number of internal filter parameters
then we can resample with linear interpolation of the filter parameters.
\end{itemize}
The disadvantage of our approach is that we cannot write
something simple like
\code{lowpass (sine controlRate) (input sampleRate)}
anymore, but with \haskell{}'s \keyword{type class} mechanism
we let the \haskell{} compiler choose the right filter
for a filter parameter type
and thus come close to the above concise expression.

\subsection{Vectorisation}
\seclabel{vectorisation}

Modern processors have vector units like
AltiVec in PowerPC,
SSE and MMX in X86 processors and Neon in ARM.
These vector units are capable of performing an operation
on multiple numbers at once.
E.g.\ a processor equipped with the SSE1 extension can perform
4 single precision floating point multiplications with one instruction.
In contrast to pure single-instruction-multiple-data (SIMD) architectures
like todays GPUs
these vector units also support rearrangement of the vector components.
Fortunately \abbr{LLVM} supports vector units through a uniform interface
and moreover it allows us to directly call processor specific instructions.
This way we have implemented various optimisations for SSE.

We have checked the use of vector units in four ways:
\begin{itemize}
\item for parallel processing
like filter banks,
processing of stereo signals
or in multi-oscillators generating a chorus effect,
a mixture of harmonics or a chord,
\item for serial processing
by dividing a signal into chunks of the length of the native vector size,
\item for pipeline processing
like in an allpass cascade or
an implementation of \person{Butterworth} or \person{Chebyshev} filters
by a cascade of second order filters \cite{hamming1989digitalfilter},
\item for internal repetitive operations
like dot products in filters.
\end{itemize}
We prefer to choose one way for all involved processes
in order to avoid expensive rearrangement of the data.
Our experiments show that the possibility for pipelining is rare
and moreover pipelining introduces a delay.
This is especially a problem for our model of causal processes.
The optimisation of internal operations is mostly restricted to filters.
Parallel vectorisation is possible only in cases
where we do the same operations in parallel
and the maximum possible speedup can only be achieved
if the number of parallel channels is a multiple of the native vector size.
Serial vectorisation is almost always possible
due to the nature of most basic signal processes.
However it requires that the user accepts
a reduction of the time resolution in cutting operations
by the factor of the size of a native vector.
It turns out that even recursive processes can be vectorised
but we may reduce the number of computations from $n$ to $\log_2 n$ only.

Coincidentally the loops for scalar computation
are often the same as the ones for
parallel vectorisation and serial vectorisation.
Only the parameters are different.
The serially vectorised counterpart
of an oscillator with initial phase $p$ and frequency $f$
for vectors of size 4
is an oscillator with initial phases
$(p, \fraction(p+f), \fraction(p+2f), \fraction(p+3f))$
and frequencies $(4f, 4f, 4f, 4f)$.
In both cases the computation of the next phases
consists of an addition and the projection into the interval $[0,1)$.
Thus serial vectorisation is just parallel vectorisation
of processes that run at the same lower rate but at interleaved phases.
We like to refer to \figref{goal-assembly} again
that shows real assembly code for a serial vectorisation.

Many generators
(linear ramps, exponential curves,
polynomial curves implemented by difference schemes or in a direct way,
noise by linear congruences)
and stateless causal processes
(mixing, ring modulation,
convolution (``non-recursive filters''),
phase modulation in oscillators,
mapping from oscillator phase to wave value,
distortion)
can be vectorised in this style.

The vectorisation of stateful causal processes is different,
if it is possible at all.
Consider an oscillator with a frequency modulated at sample rate.
Computing the phases means computing the cumulative sum
followed by a parallel computation of the fraction of all interim sums.
E.g.\ the cumulative sum $d$ of an 8-element vector $a$ can be computed by
\begin{eqnarray}
b &=& a + a\shiftup1 \\
c &=& b + b\shiftup2 \\
d &=& c + c\shiftup4
\end{eqnarray}
where $x\shiftup n$ denotes shifting the vector $x$ by $n$ components upwards,
filling the least components with zeros.
The same way we can write a first order filter with feedback factor $k$.
\begin{eqnarray}
b &=& a + k  \cdot a\shiftup1 \\
c &=& b + k^2\cdot b\shiftup2 \\
d &=& c + k^4\cdot c\shiftup4
\end{eqnarray}
We can express this by the $z$-transformation of that filter:
\[
\frac{1}{1-k\cdot z^{-1}}
 = \frac{1+k\cdot z^{-1}}{1-k^2\cdot z^{-2}}
 = \frac{(1+k\cdot z^{-1})\cdot(1+k^2\cdot z^{-2})}{1-k^4\cdot z^{-4}}
\]
Generally by extending a polynomial fraction
with the alternating polynomial of the denominator
we eliminate all monomials with odd exponent in the denominator.
This way we can decompose a purely recursive filter of any order
into a short-term non-recursive filter
and a long-term recursive filter.
For a second order filter we obtain
\[
\frac{1}{1-a\cdot z^{-1}+b\cdot z^{-2}}
 = \frac{1+a\cdot z^{-1}+b\cdot z^{-2}}
        {1-(a^2-2\cdot b)\cdot z^{-2} + b^2\cdot z^{-4}}
\]
and we repeat that extension until the non-recursive filter mask
reaches the native vector size,
that is usually a power of two.
In a real application we must cope with varying filter parameters.
To this end we had to adjust the general principle
in order to get the same results for scalar and vector implementations
that are controlled at ``vector rate''
(sample rate divided by native vector size).

\subsection{Parameters at Runtime}
\seclabel{parameter}

So far we have only considered signal generators with parameters
that are hard-wired into the machine loop.
This means however that when rendering a song
we need to recompile the signal generator for every tone
according to its pitch, its velocity and the gate signal.
In order to overcome this
we let the user define a record type \code{p}
that contains all parameters he wishes to control.
The generator type becomes \code{Generator p a}
and e.g.\ the type of an exponential curve generator becomes
\begin{verbatim}
exponential ::
   (p -> Float) -> (p -> Float) ->
   Generator p (Value Float)
\end{verbatim}
where the parameters are functions that get a value from the record \code{p}
(\keyword{record field selectors}).
The function
\begin{verbatim}
compile ::
   Generator p (Value a) ->
   IO (p -> StorableVector a)
\end{verbatim}
compiles the generator via \abbr{LLVM}
and returns a function that depends on the parameter set of type \code{p}.
%
%
In our implementation we distinguish between constant parameters
and open parameters
and hard-wire constant parameters into the machine loop.

\section{Related Work}

Our goal is to make use of the elegance of \haskell{} programming
for signal processing.
Our work is driven by the experience that today compiled \haskell{} code
cannot compete with traditional signal processing packages written in C.
There has been a lot of progress in recent years,
most notably the improved support for arrays without overhead,
the elimination of temporary arrays (\keyword{fusion})
and the Data-Parallel Haskell project \cite{peyton-jones2008dph}
that aims at utilising multiple cores of modern processors
for array oriented data processing.
However there is still a considerable gap in performance
between idiomatic \haskell{} code and idiomatic C code.
A recent development is an \abbr{LLVM}-backend
for the \abbreviation{GHC}{Glasgow Haskell Compiler}.
that adds all of the low-level optimisations of \abbr{LLVM} to \abbr{GHC}.
However we still need some tuning of the high-level optimisation
and a support for processor vector types
in order to catch up with our \abbr{EDSL} method.


In \secref{background} we gave some general thoughts
about possible designs of signal processing languages.
Actually for many combinations of features we find instances:
The two well-established packages
CSound \cite{vercoe2004csound}
and SuperCollider \cite{mccartney1996supercollider}
are domain specific untyped languages
that process data in a chunky manner.
This implies that they have no problem
with sharing signals between signal processors
but they support feedback with short delay only by small buffers (slow)
or by custom plugins (more development effort).
Both packages support three rates:
note rate, control rate and sample rate
in order to reduce expensive computations
of internal (filter) parameters.
With the \haskell{} wrappers
\cite{hudak1996haskore,drape2009hsc3-0.7}
it is already possible to control these programs
as if they were part of \haskell{},
but it is not possible to exchange audio streams with them in real-time.
This shortcoming is resolved with our approach.

Another special purpose language is ChucK \cite{wang2004chuck}.
Distinguishing features of ChucK
are the generalisation to many different rates
and the possibility of programming while the program is running,
that is while the sound is playing.
As explained in \secref{internal-parameter}
we can already cope with control signals at different rates,
however the management of sample rates at all could be better
if it was integrated in our framework for physical dimensions.
Since the \haskell{} systems Hugs and \abbr{GHC}
both have a fine interactive mode,
\haskell{} can in principle also be used for live coding.
However it still requires better support
by LLVM (shared libraries) and by our implementation.

Efficient short-delay feedback written in a declarative manner
can probably only be achieved
by compiling signal processes to a machine loop.
This is the approach implemented by
the Structured Audio Orchestra Language of MPEG-4
\cite{iso1999saol}
and
Faust
\cite{orlarey2004faust}.
Faust started as compiler to the C++ programming language,
but it does now also support LLVM.
Its block diagram model very much resembles
\haskell{}'s arrows (\secref{causal-process}).
A difference is that Faust's combinators
contain more automatisms
which on the one hand simplifies binding of signal processors
and on the other hand means that errors in connections cannot be spotted locally.

Before our project the compiling approach
embedded in a general purpose language was chosen by
Common Lisp Music \cite{schottstaedt2009clm},
Lua-AV \cite{smith2007luaav},
and Feldspar (\haskell)
\cite{technology2009feldspar-language-0.1}.

Of all listed languages only ChucK and \haskell{}
are strongly and statically typed,
and thus provide an extra layer of safety.
We like to count Faust as being weakly typed
since it provides only one integer and one floating point type.

\section{Benchmarks}
\seclabel{benchmark}

We like to put the performance of our implementation
in the context of existing signal processing packages.
To this end we compare it with the well-established packages
CSound 5.10 and SuperCollider 3 on an X86 machine with SSSE3
in \tabref{benchmarks}.
The code for the examples can be found in the
\href{http://code.haskell.org/synthesizer/llvm/dafx2010/}{dafx2010} directory
of our repository \cite{thielemann2010synthillvm-darcs}.
SuperCollider is designed as real-time server,
but we run it in non-real-time mode.
In order to increase relative time measuring precision
we chose a large number of generated samples,
namely $200\cdot 44100$ that is about 10 million samples.
We generate single precision float samples
since this is the native format of both packages.
We measure the ``user time'' with UNIX's \code{time},
thus the time for writing to the disk is ignored.

The columns ``scalar'' and ``vector'' refer to our implementation
with scalar and vector operations up to SSSE3, respectively.
As far as we know
both CSound and SuperCollider do not use the X86 vector unit
in the tested versions.
Compilation and optimisation by LLVM-2.6
is part of the execution time for our implementation,
but it is really not noticeable.

``Saw'' denotes a sawtooth oscillator with constant frequency
and no anti-aliasing.
That is, for SuperCollider we use \code{lfSaw}
instead of the band-limited \code{saw}.
It is certainly the most simple waveform we can generate.
In CSound the particular waveform does not matter
since the oscillator reads it from a table.
``Ping'' is the same sawtooth enveloped in an exponential curve
and shall check whether it helps
to fuse different signal generators into one loop.
We have chosen the half-life large enough
in order to not run into denormalised numbers.
Their handling is expensive,
but it may also be avoided by the round-denormalised-to-zero mode.
``Chord'' is a mixture of four sawtooth tones at different frequencies
and shall demonstrate whether parallel vectorisation pays off.
The number of tones is small enough to hold all loop variables in SSE registers.
``Chordchorus'' is a mixture of four chorus oscillators,
each consisting of four plain sawtooth oscillators,
that is a total of 16 oscillators.
Here the number of SSE registers is exceeded.
``Butterworth'' is a filter sweep of a 10th order \person{Butterworth} lowpass filter
applied to white noise
where the control rate is a 100th of the sample rate.
Since SuperCollider does not provide a higher order \person{Butterworth} filter
we use a cascade of 5 second order \person{Butterworth} filters (\code{lpf}).
The result is different but the speed should be comparable.
This example shall demonstrate that our vectorised implementation
of second order filters actually gives a little increase in performance.
The noise however is vectorised, too.
``Allpass'' is a phaser implemented as causal process using an allpass cascade.
It is applied to the expensive ``Butterworth'' generator
in order to show the savings of sharing the original and the allpass filtered signal.
However, for SuperCollider we have chosen a simple delay line
because of the lack of a first order allpass.
``Karplus-Strong'' is a variant of the according algorithm
using a feedback with delay of 100 samples and a first order lowpass.
\begin{table}
\newcommand\perftime[3]{#1}
\begin{center}
\begin{tabular}{|l|r|r|r|r|}
\hline
 & CSound & SuperCollider & scalar & vector \\
\hline
%
%
saw
 & \perftime{0.24}{0.242}{0.030}
 & \perftime{0.09}{0.094}{0.017}
 & \perftime{0.08}{0.082}{0.019}
 & \perftime{0.02}{0.023}{0.006}
 \\
chord
 & \perftime{0.54}{0.540}{0.039}
 & \perftime{0.23}{0.228}{0.023}
 & \perftime{0.13}{0.130}{0.010}
 & \perftime{0.04}{0.038}{0.007}
 \\
chordchorus
 & \perftime{1.54}{1.539}{0.081}
 & \perftime{0.77}{0.765}{0.013}
 & \perftime{0.49}{0.495}{0.040}
 & \perftime{0.13}{0.134}{0.010}
 \\
ping
 & \perftime{0.25}{0.246}{0.031}
 & \perftime{0.11}{0.113}{0.008}
 & \perftime{0.10}{0.101}{0.020}
 & \perftime{0.03}{0.026}{0.007}
 \\
butterworth
 & \perftime{1.22}{1.222}{0.059}
 & \perftime{0.45}{0.454}{0.013}
 & \perftime{0.74}{0.738}{0.039}
 & \perftime{0.60}{0.604}{0.032}
 \\
allpass
 & \perftime{1.54}{1.535}{0.051}
 & \perftime{0.58}{0.582}{0.030}
 & \perftime{1.07}{1.066}{0.057}
 & \perftime{0.84}{0.835}{0.042}
 \\
karplus
 & \perftime{1.64}{1.635}{0.060}
 & \perftime{0.34}{0.340}{0.042}
 & \perftime{0.08}{0.081}{0.009}
 & \perftime{0.06}{0.057}{0.009}
 \\
\hline
\end{tabular}
\end{center}
\vspace{-1.7\baselineskip}
\tabcaption{benchmarks}{
Benchmarks for computing 8820000 (200 seconds at 44100 Hz)
single precision floating point samples.
Computing times are given in seconds.
}
\end{table}

\section{Conclusions and further work}

The speed numbers of our implementation are excellent,
yet the generating \haskell{} code looks idiomatic.
The next step is the integration of the current low-level implementation
into our existing framework for signal processing
that works with real physical quantities
and statically checked physical dimensions.
There is also a lot of room for automated optimisations by \abbr{GHC} rules,
be it for vectorisation or
for reduction of redundant computations of $\fraction$.

We hope that \abbr{LLVM} supports GPUs in the future
and thus makes them accessible by our framework.
However this will certainly need a concerted effort for the \abbr{LLVM} developers
and it also requires an adapted design of our vectorisation,
since GPUs have no notion of a vector and thus cannot shuffle vector elements.
Actually \abbr{LLVM} can already generate C code,
but we have not checked whether this is compatible with CUDA.

\section{Acknowledgments}
I like to thank the careful proofreaders
J\"{u}rgen Donath, Josef Svenningsson, Emil Axelsson,
Johannes Waldmann, Erik de Castro Lopo, Stefan Kersten,
Graham Wakefield and Wolfgang Jeltsch
for their valuable suggestions.

\bibliographystyle{IEEEbib}
\bibliography{audio,haskell,thielemann,hackage,literature}

\end{document}